\documentclass[11pt,twoside]{article}

\usepackage{asp2004}
\usepackage{epsf}
\usepackage{psfig}
\usepackage{lscape}
\usepackage{graphicx}
\usepackage{array}
\usepackage{amsmath}
\usepackage{amsfonts}

\begin{document}
\title{Magnetohydrodynamic Simulation of Solar Supergranulation}   
\author{S. D. Ustyugov}   
\affil{Keldysh Institute of Applied Mathematics, 4, Miusskaya sq., Moscow,
Russia}    

\begin{abstract} 
Three-dimensional magnetohydrodynamical large eddy simulations
of solar surface convection using realistic model physics is conducted.
The effects of magnetic fields on thermal structure of convective motions
into radiative layers, the range of convection cell sizes and penetration depths
of convection is investigated. We simulate a some portion of the solar photosphere
and the upper layers of the convection zone, a region extending 30 x 30 Mm
horizontally from 0 Mm down to 18 Mm below the visible surface.
We solve equations of the fully compressible radiation magnetohydrodynamics
with dynamical viscosity and gravity.  For numerical simulation we use:
1)realistic initial model of Sun  and equation of state and opacities of
stellar matter, 2) high order conservative TVD scheme for solution
magnetohydrodynamics, 3) diffusion approximation for solution radiative transfer
4) calculation dynamical viscosity from subgrid scale modelling.
Simulations are conducted on horizontal uniform grid of 320 x 320
and with 144 nonuniformly spaced vertical grid points on the 128 processors
of supercomputer MBC-1500 with distributed memory multiprocessors 
in Russian Academy of Sciences. 

\end{abstract}



\section{Introduction}
Convection near solar surface has strongly non-local and dynamical
character. Hence numerical simulation provide useful information on
the structure spatial scales by convection and help to construct
consistent models of the physical processes underlying the observed
solar phenomena. We investigate effects of compressibility and weak
of magnetic field on formation non-local structure of convection 
using realistic physics and conservative TVD numerical scheme of Godunov type. 
The previous simulations were confined by small computational domain and studied
processes on scales order size of granulation [\citet{stein06}].
In order to investigate collective interaction of convective modes
different scales and process of formation of supergranulation
we conducted calculation in three dimensional computational box by
size 30 Mm in horizontal direction and by size 18 Mm in vertical
direction.

\section{Numerical method}

We take distribution of the main thermodynamic variables  by radius
due to Standard Solar Model [\citet{christ03}] with parameters
$(X,Z,\alpha)=(0.7385,0.0181,2.02)$, where $X$ and $Y$ are hydrogen
and helium abundance by mass, and $\alpha$ is the ratio of mixing
length to pressure scale height in convection region. We use
OPAL opacities and equations of state for solar matter [\citet{rogers96}].

We solve fully compressible nonideal magnetohydrodynamics equations:

$$\frac{\partial \rho}{\partial t}+ \nabla \cdot \rho \vec v = 0$$
$$\frac{\partial \rho \vec v}{\partial t}+ \nabla \cdot \left[\rho \vec v \vec v
+\left( P +  \frac {B^2}{8 \pi}\right)I - \frac {\vec B \cdot \vec B}{4 \pi}\right]
= \rho \vec g + \nabla \cdot \tau$$
$$\frac{\partial E}{\partial t}+ \nabla \cdot \left[\vec v
\left(E + P +  \frac {B^2}{8 \pi}\right) - \frac {\vec B \left(\vec v \cdot \vec B\right)}{4 \pi}\right]
$$
$$= \frac {1}{4 \pi}\nabla \cdot \left(\vec B \times \eta \nabla \times \vec B\right)
+\nabla \cdot \left(\vec v \cdot \tau \right) + \rho \left(\vec g \cdot \vec v \right) + Q_{rad}$$
$$\frac{\partial \vec B}{\partial t}+ \nabla \cdot \left(\vec v \vec B - \vec B \vec v\right)=
- \nabla \times \left(\eta \nabla \times \vec B \right)$$

where $E=e+\rho v^2/2+B^2/8 \pi $ is the total energy, $Q_{rad}$ is the
energy transferred by radiation and $\tau$ is viscous stress tensor.

We assume that small scales are independent of resolved scales
(Large Eddy Simulations) and rate dissipation is defined from
buoyancy and shear production terms [\citet{canuto94}].
The numerical method that we used was an explicit Godunov-type conservative
TVD difference scheme [\citet{yee90}]

$$
U^{n+1}_{i,j,k}=U^n_{i,j,k}-\Delta tL(U^n_{i,j,k}),
$$

where $\Delta t=t^{n+1}-t^n$ and operator $L$ is

$$
\begin{array}{r}
L(U_{i,j,k})=\cfrac {\tilde F_{i+1/2,j,k}-\tilde F_{i-1/2,j,k}}{\Delta x_i}
+ \cfrac {\tilde G_{i,j+1/2,k}-\tilde G_{i,j-1/2,k}}{\Delta y_j}\\
+ \cfrac {\tilde H_{i,j,k+1/2}-\tilde H_{i,j,k-1/2}}{\Delta z_k}+S_{i,j,k}
\end{array}
$$

Flux along each direction, for example x,  was defined
by local-characteristic method  as follows
$$
\tilde F_{i+1/2,j,k}=\frac {1}{2}\left[F_{i,j,k}+F_{i+1,j,k}+
R_{i+1/2}W_{i+1/2}\right]
$$

where $R_{i+1/2}$ is matrix whose columns are right eigenvectors of
$\partial F/\partial U$ evaluated at generalized Roe average for real
gases of $U_{i,j,k}$ and $U_{i+1,j,k}$. The $W_{i+1/2}$ is the matrix of
numerical dissipation. Term $S_{i,j,k}$ is accounted effect of
gravitation forces and radiation.

The one step of time integration is defined by Runge-Kutta
method [\citet{shu88}] as

\newpage

$$
U^{(1)}=U^n+\Delta tL(U^n)
$$
$$
U^{(2)}=\frac {3}{4}U^n+\frac {1}{4}U^{(1)}+\frac {1}{4}\Delta
tL(U^{(1)})
$$
$$
U^{n+1}=\frac {1}{3}U^n+\frac {2}{3}U^{(2)}+\frac {2}{3}\Delta
tL(U^{(2)})
$$

The scheme is second order by space and time. For approximation
viscous terms we used central differences. For evaluation
radiative term in energy equation we used the diffusion
approximation

$$
Q_{rad}= \nabla \cdot \left[\frac{4acT^3}{3k\rho} \nabla T \right]
$$

We use uniform grid in x-y directions and nonuniform grid
in vertical z one. We apply periodic boundary condition
in horizontal planes and choose on the top and bottom as follows
$$v_{z,k}=-v_{z,k\pm1}, v_{x,k}=v_{x,k\pm1}, v_{y,k}=v_{y,k\pm1}$$
$$dp/dz=\rho g_{z}, p=p(\rho), e=const$$
$$B_{x}=B_{y}=0, dB_{z}/dz=0$$

In initial moment magnetic field equal 50 G and has just one vertical
component. For the magnetic diffusitvity we take constant value
$\eta = 1.1\times10^{11}$cm${}^2$sec${}^{-1}$

\section{Results}

On the figures 1-4 results of development convection after 12 solar hr MHD 
numerical simulation are shown. We founded that magnetic field concentrate 
on the boundary of convective cells in forms magnetic flux and sheets. 
Diverging of convective flows from centre supergranular expell weak magnetic 
field on the edges of convective cell. Average size of supergranular celss 
is 10-20Mm with lifetime about 8-10 hours. We aplly procedure averaging 
by interval time two hour and find value of velocity from centre 
supergranular equal to 1-1.5 km/sec. In places action of strong magnetic 
field with strength 700-900 G we observe effect of suppression of convection 
and decreasing of fluctuation of temperature. Magnetic pressure in regions 
concentration of magnetic flux prevent inflow of matter. Transfer of 
radiation energy in these places is suppressed. We founded from simulation 
that maximum of value of magnetic field in computational domain equal to 1300 G. 

Inside of supergranular we have usual picture of evolution of convection
on scale of granulation with average sizes of cells about 1-2Mm and 
lifetimes about 10 minutes. Here we see wider upflows of warm, 
low density, and entropy neutral matter and downflows of cold, converging 
into filamentary structures, dense material. We observe continuous picture 
formation and destruction of granules. Granules with highest pressure 
grow and push matter against neighboring granules, that then shrink 
and disappear. Ascending flow increases pressure in center of granule 
and upflowing fluids decelerates motion. This process reduce heat 
transport to surface and allow material above the granule to cool, 
become denser, and by action gravity to move down. We observe formation 
new cold intergranule lane splitting the original granule.

From figure 4 we see distinctly existence three different regions 
development of convection. In near of solar surface to the depth 4 Mm 
we founded zone of turbulent convection. In this region cold blobs 
of matter move down with velocity in maximum about 4 km/sec with
maximum Mach number equal 1. The downdrafts has different and 
very complicate vertical structure. Some ones travel small distance 
from surface and become weak enough to be broken up by the surrounding 
fluid motion. Other ones conserve motion with high velocity and 
move on distance about 6 Mm. We observe that different nature such 
behavior is due to initial condition of formation downdrafts. 
In place of confluence of convective cells to one point more energy 
is released.

In region from 5Mm to 8Mm of depth we reveal more quiet character
of convective flow than in turbulent zone. Below 8Mm we see 
clear separate large scale density fluctuations and streaming flow 
of matter similar jets with average velocity about 1km/sec. 
The magnetic field  in these places has values about 300G. 
Distance beetween different narrow jets gives size of supergranular 
cells. Places of intersections of path jets with horizontal plane 
are vertices of huge convective cells. From distribution iso-surface 
of magnetic field with strength 500 G we have discovered that pumping 
of magnetic field occupy more part of all computational domain up to 
bottom boundary.  

Magnetic field near solar surface in our numerical simulation
has very complicate structure. Inside supergranular cells we see
formation, growth and evolution with time many arising to surface 
loops of magnetic field. There is places in vertices of huge cells 
with vorticity motion that provide quick rate magnetic helicity 
transport across solar photosphere(Fig.5,6). On the boundary supergranular 
cell magnetic field has just vertical component. In these parts we 
observe quick changing sign and big values of current helicity(Fig.7).
 
\section{Acknowledgments}

I would like to thank Gary Zank and Nikolai Pogorelov 
from Riverside University for financial support for
participation me in conference Astronum-2006.






\begin{figure}[h+]
  \includegraphics[scale=0.63]{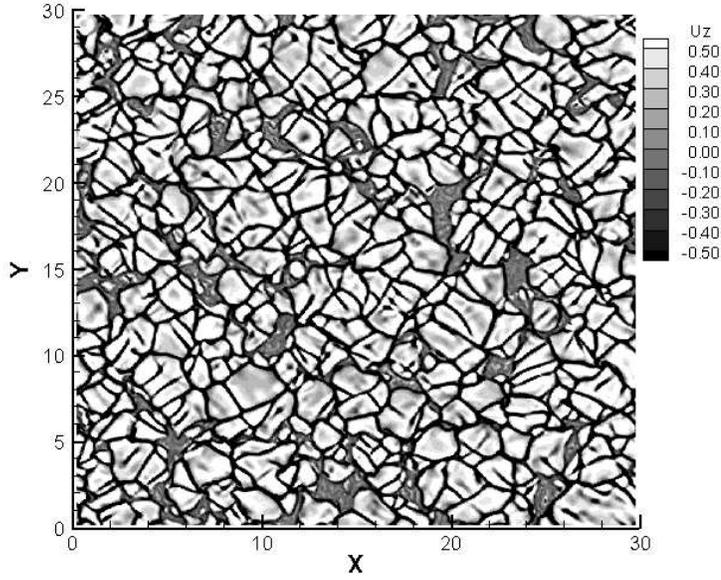}
  \caption{Image of contours of the vertical velocity in horizontal plane 
near solar surface.}
\end{figure}

\begin{figure}[h+]
  \includegraphics[scale=0.63]{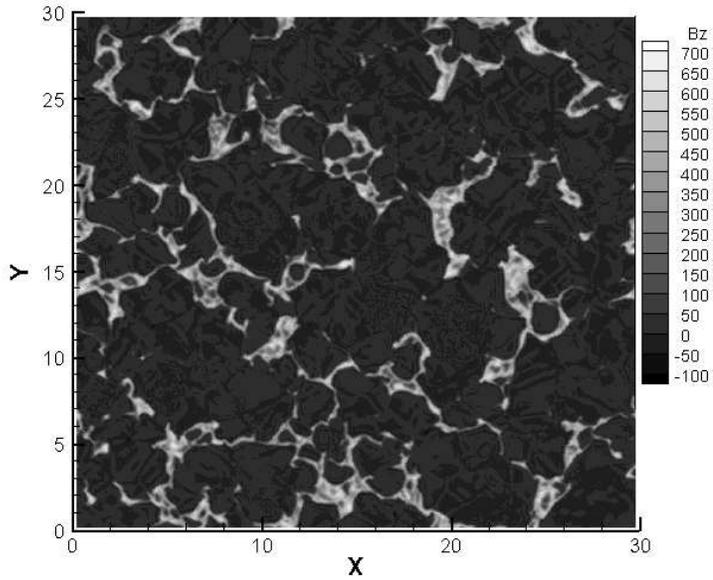}
  \caption{Image of contours of the vertical component of magnetic field at the same 
plane as that for Figure 1.}
\end{figure}

\begin{figure}[h+]
  \includegraphics[scale=0.63]{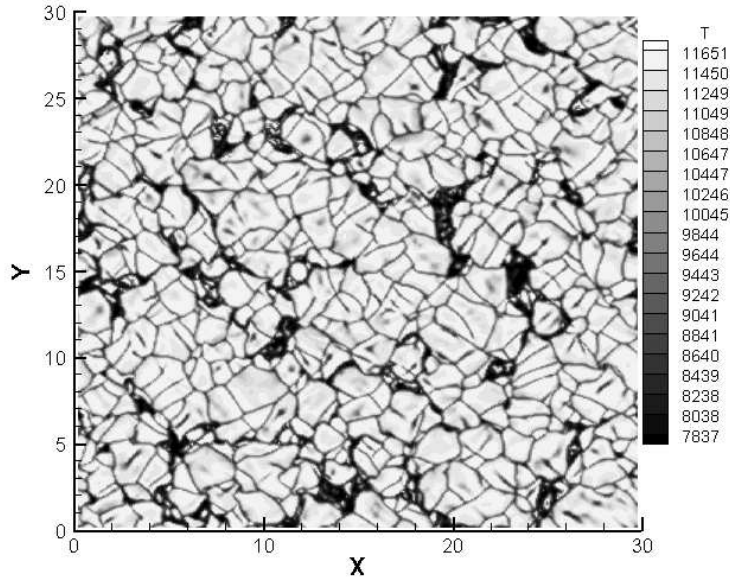}
  \caption{Image of contours of the temperature in the horizontal plane near solar surface}
\end{figure}

\begin{figure}[h+]
  \includegraphics[scale=0.63]{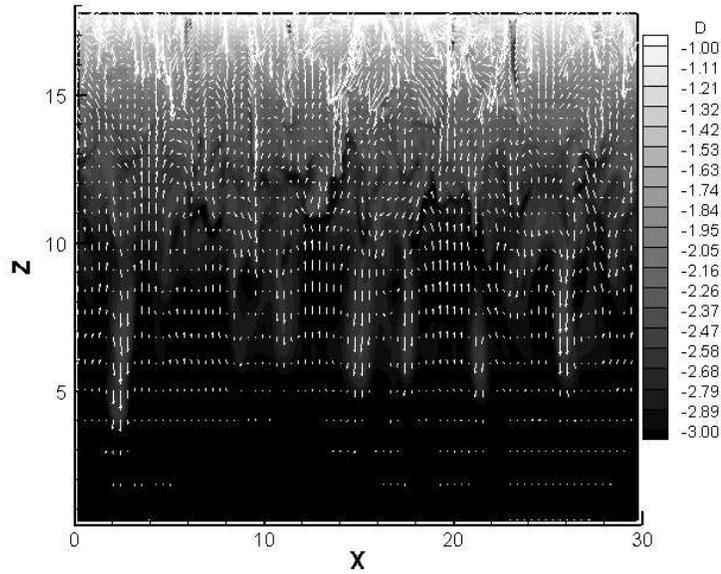}
  \caption{Image of contours of the fluctuations density and field of velocity in vertical plane.}
\end{figure}

\begin{figure}[h+]
  \includegraphics[scale=0.45]{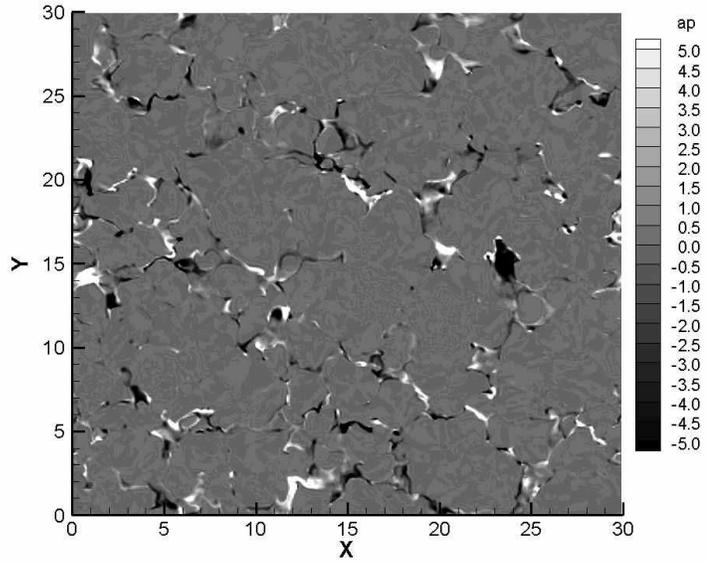}
  \caption{Image of contours of the magnetic helicity transport rate in horizontal plane 
near solar surface as that for Figure 1.}
\end{figure}

\begin{figure}[h+]
  \includegraphics[scale=0.45]{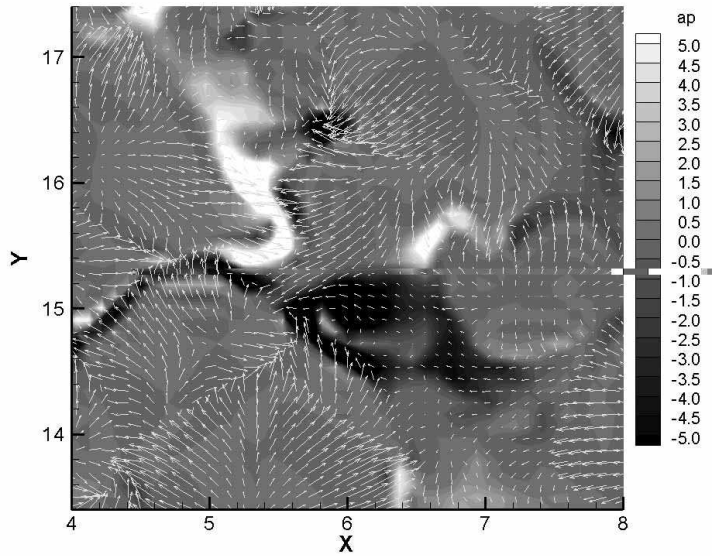}
  \caption{Image of contours of the magnetic helicity transport rate and field of velocity 
in small part computational domain at the same plane as that for Figure 5.}
\end{figure}

\begin{figure}[h+]
  \includegraphics[scale=0.45]{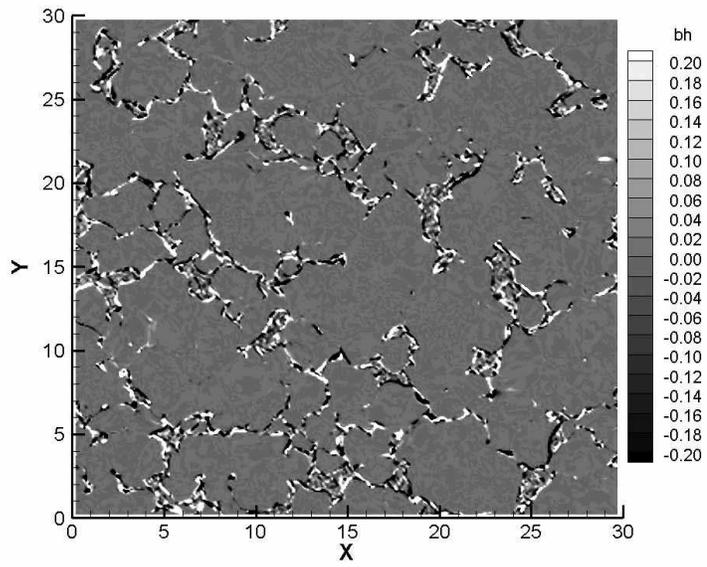}
  \caption{Image of contours of the current helicity at the same plane as that for Figure 1.}
\end{figure}

\end{document}